\makeatletter
\declare@file@substitution{revtex4-1.cls}{revtex4-2.cls}
\makeatother
\documentclass[twocolumn]{aastex631}
\usepackage[normalem]{ulem}
\usepackage{graphicx}
\usepackage{dcolumn}
\usepackage{bm}
\usepackage{color}
\usepackage{array}
\usepackage{multirow}
\usepackage{amsmath}
\usepackage{bm}
\usepackage{subfigure}

\usepackage{float}
\usepackage{hyperref}

\begin{document}

\title{Cosmological constraints using high-redshift galaxies at $5 < z < 14$ from JWST}

\author{L. Huang}
 \affiliation{College of Science, Jiujiang University, Jiujiang 332000, People's Republic of China;huanglong20122021@163.com}

\author{ P. Zong}
 \affiliation{School of Physics and Electronic Engineering, Fuyang Normal University, Fuyang 236000, People's Republic of China.}

 \author{ Z. X. Chang}
 \affiliation{College of Mathematical and Physical Sciences, HanDan University, Handan 056000, People's Republic of China.}

\author{ N. Chang}
\affiliation{Xinjiang Astronomical Observatory, Chinese Academy of Sciences, \\
Urumqi 830011, People's Republic of China.}
\affiliation{Key Laboratory of Radio Astronomy, Chinese Academy of Sciences, \\
Nanjing, 210008, People's Republic of China.}

 \author{ F. He}
\affiliation{Institute of Physics and Electronic Science, Hunan University of Science and Technology, \\
Xiangtan, Hunan 411201, People's Republic of China.}

\begin{abstract}

We use the star formation rate-stellar mass relation ($SFR - {M_ * }$) of galaxies to measure their luminosity distances as well as estimate cosmological parameters. We compile a sample of 341 high-redshift galaxies at $5 < z < 14$ and an additional 51 galaxies at $1.8 < z < 3.5$ from the JWST observations, which can be used to investigate the correlation between the star formation rate and stellar mass, and determine their cosmological distance. The relation $SFR \propto {\kern 1pt} {({M_ * }/{M_ \odot })^\gamma }$ can be applied to check the nonlinear correlation between the star formation rate and stellar mass ${M_ * }$ of galaxies and give their distance. Additionally, it also can be used to test if the $SFR - {M_ * }$ relation depends on redshift, and the data suggest that the $SFR $ is positively correlated with stellar mass, the $SFR - {M_ * }$ relation of the high-redshift galaxies has a redshift dependence. Finally, we combine the galaxy sample with Type Ia supernova (SNIa) data from the Pantheon compilation to test the property of dark energy, specifically addressing whether its density deviates from the constant, and give the statistical analysis results.

\end{abstract}


\section{Introduction} \label{Sec:1}

A great number of galaxies data have been obtained and used to investigate the relationship between the star formation rate (SFR) and stellar mass (${M_ * }$) (known as main sequence), which involves that how galaxies build up their stellar mass in galaxy formation.

Brinchmann (2004) used the spectroscopic data of low-redshift galaxies from the Sloan Digital Sky Survey (SDSS) to investigate the correlation between the stellar mass and the SFR \citep{Brinchmann2004}. Subsequently, more galaxies data are used to verify the star formation rate-stellar mass relation ($SFR - {M_ * }$) \citep{Daddi2007, Elbaz2007, Noeske2007, Dave2008}. The $SFR - {M_ * }$ relation plays an important role in constraining theoretical models of galaxy evolution and is one of the basic relations describing the evolution of galaxies.

 The different research work for $SFR \propto {\kern 1pt} {({M_ * }/{M_ \odot })^\gamma }$ relation have different slope $\gamma$, there are  several main reasons for this phenomenon. $(1)$ The differences in study sample selectivity, such as magnitude-limited sample, mass-limited sample, and multi-band color sample selection; (2) Different criteria for selecting star-forming galaxies (SFGs); (3) The calculation method for star formation rate of galaxies are inconsistent; (4) Differences in measurement techniques of stellar mass.

On the other hand, large amounts of galaxy data indicate that there is a redshift evolution of the star formation rate-stellar mass relation \citep{Koyama2013, Whitaker2012}.  It shows that high redshift galaxies might form stars faster at a certain mass. The evolution of $SFR - {M_ * }$ relation of galaxies with the redshift also can be used to represent the history of star formation in the universe.

In this paper, we introduce the source of data used in Section \ref{Sec:2}, including the star formation rate and stellar mass of 341 high-redshift galaxies at $5 < z < 14$, and 51 others at $1.8 < z < 3.5$. In Section \ref{Sec:3}, we employ the nonlinear relation $SFR \propto {\kern 1pt} {({M_ * }/{M_ \odot })^\gamma }$ to check the correlation between the star formation rate and stellar mass ${M_ * }$ of galaxies, and give their cosmological luminosity distance. In Section \ref{Sec:4}, we apply a combination of high redshift galaxies and Type Ia supernova (SNIa) Pantheon to reconstruct the dark energy equation of state $w(z)$, which can be used to test the nature of dark energy concerning whether or not its density deviates from the constant. In Section \ref{Sec:5}, we summarize the paper.

\section[ data used]{ data used} \label{Sec:2}

Modern optical including Sloan Digital Sky Survey(SDSS) \citep{Lyke2020, Ahumada2020}, Hubble Space Telescope (HST) \citep{Skelton2014, Bouwens2015}, and James Webb Space Telescope (JWST) provide the optical spectra from UV to infrared for a large amount of galaxies \citep{naidu2022two, Bakx2023}, especially for the JWST that observes a lower frequency range, from long-wavelength visible light (red) through mid-infrared (0.6-28.3 $\mu m$ ), which can provide rest-frame UV/optical spectra of high-redshift galaxies when their visible emissions have shifted into the infrared. The high-redshift galaxies can be used to investigate the formation of early galactic structure.

\citet{Li2023} presented a sample of 51 galaxies at $z = 2-3$ from the A2744 and SMACS J0723-3732 (SMACS 0723) galaxy cluster fields by deep JWST/NIRISS imaging and slitless grism spectroscopic observations, where 44 are in A2744, and seven are in SMACS 0723. Their star formation rate (SFR) can be measured by $H\alpha $ or $H\beta $ emission lines. The galaxy stellar masses are derived by  \citet{Chabrier2003} initial mass function (IMF) combined with stellar mass-to-light ratios (M/L) model after correcting the magnitudes in all bands by the lensing magnification factor $\mu $  \citep{Wambsganss1998}, $\mu $ can be calculated by the entire cluster mass distribution and angular positions of the sources \citep{Mahler2018, Golubchik2022}.

\citet{Morishita2024} also obtained 341 high-redshift galaxies at $5 < z < 14$, which are based on nine public deep extragalactic fields from JWST Cycle 1. The SFR for high-redshift galaxies without $H\alpha $ and $H\beta $ coverage can be calculated by the corrected UV luminosity  for lensing magnification factor $\mu $ and dust attenuation, and UV flux can be fitted from the extracted spectrum after calibrating the raw images by Grizli \footnote{https://github.com/gbrammer/grizli/} \citep{Morishita2021, Bergamini2023, Morishita2024}.

Therefore, we consider to use a combined data from  \citet{Li2023} and \citet{Morishita2024}, this joint sample of 392 galaxies can be applied to check the correlation between the star formation rate and stellar mass ${M_ * }$ of galaxies and calculate their luminosity distance. The data can be obtained from dol:  \href{https://archive.stsci.edu/doi/resolve/resolve.html?doi=10.17909/q8cd-2q22}{10.17909/q8cd-2q22}, \href{https://archive.stsci.edu/doi/resolve/resolve.html?doi=10.17909/91zv-yg35}{10.17909/91zv-yg35}, and \href{https://archive.stsci.edu/doi/resolve/resolve.html?doi=10.17909/12rr-2a67}{10.17909/12rr-2a67}.

\section{Parameter constraints of $SFR - {M_ * }$ relation of galaxies from JWST}\label{Sec:3}
\subsection{Insights from scatter plots}
For the galaxies with a redshift range of $1.8 < z < 3.5$, the \citet{Kennicutt1998} SFR calibration is given by \citep{Li2023}
 \begin{eqnarray}\label{eq1}
\begin{array}{l}
SFR = 4.6 \times {10^{ - 42}}\frac{{{L_{H\alpha }}}}{{erg{\kern 1pt} {\kern 1pt} {s^{ - 1}}}}[{M_ \odot }{\kern 1pt} y{r^{ - 1}}],
\end{array}
\end{eqnarray}
 where ${L_{H\alpha }}$ represents the ${H\alpha }$ luminosity, and it has been obtained from the measured emission line flux assuming $\Lambda CDM$ cosmology $({\Omega _m} = 0.3,{\kern 1pt} {\kern 1pt} {H_0} = 70{\kern 1pt} km{\kern 1pt} {\kern 1pt} {s^{ - 1}}{\kern 1pt} Mp{c^{ - 1}})$. For galaxies without ${H\alpha }$ coverage, the ${H\alpha }$ luminosity can be derived from the attenuation-corrected ${H\beta }$ flux using Grizli, and adopting Balmer decrement of $H\alpha /H\beta  = 2.86$.

 For the galaxies at $5 < z < 14$,

  \begin{eqnarray}\label{eq2}
\begin{array}{l}
SFR = 1.4 \times {10^{ - 28}}\frac{{{L_{UV }}}}{{erg{\kern 1pt} {\kern 1pt} {s^{ - 1}}}}[{M_ \odot }{\kern 1pt} y{r^{ - 1}}].
\end{array}
\end{eqnarray}

The linear formula is usually used to investigate the correlation between the SFR and stellar mass (${M_ * }$) for galaxies, which can be written as \citep{Daddi2007, Elbaz2007, Noeske2007}
 \begin{eqnarray}\label{eq3}
\begin{array}{l}
\log SFR = \beta  + \gamma {\kern 1pt} \log (\frac{{{M_ * }}}{{{M_ \odot }}}),
\end{array}
\end{eqnarray}

The above equation is equivalent to the relation $SFR{\kern 1pt}  \propto {\kern 1pt} {({M_ * }/{M_ \odot })^\gamma }$. We fit equation (\ref{eq3}) to 341 high-redshift galaxies at $5 < z < 14$ and get the residual $\Delta (\log {\kern 1pt} SFR){\kern 1pt} $ from the statistical values of $\beta$ and $\gamma$.

\begin{figure}[htpb]
\centering
\scalebox{.80}{\includegraphics[width=\linewidth,scale=1]{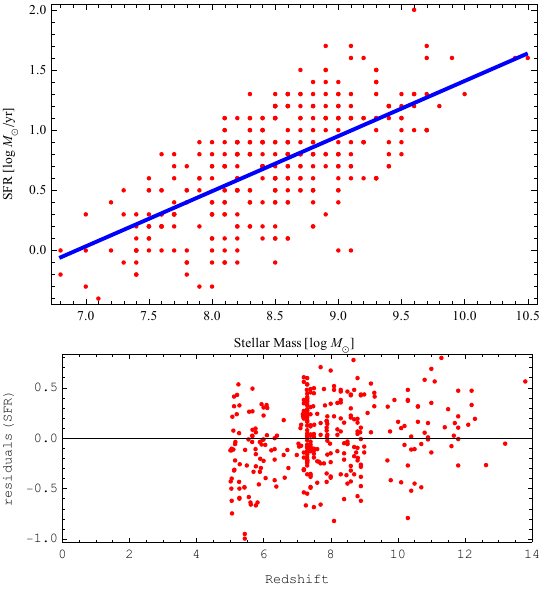}}
\caption{Plot of $\log SFR{\kern 1pt} {\kern 1pt} vs.{\kern 1pt} \log (\frac{{{M_*}}}{{{M_ \odot }}})$ (upper panel), and $\Delta (\log {\kern 1pt} {\kern 1pt} SFR){\kern 1pt} {\kern 1pt} vs.redshift$ (lower panel) for high-redshift galaxies. The blue line in the upper panel is a fitting value of $\log SFR$ from Eq. (\ref{eq3}) with the best fitting values of $\beta$ and $\gamma$. $\Delta (\log {\kern 1pt} SFR)$ represents the residuals from the subtraction of the observational data and theoretical values. In the lower panel the relationship between $\Delta (\log {\kern 1pt} SFR)$ and ${M_{UV}}$ indicates the redshift dependence of the slope $\gamma$.}
\label{fig:1}
\end{figure}

The $\log {\kern 1pt} {\kern 1pt} SFR - \log {\kern 1pt} {\kern 1pt} ({M_ * }/{M_ \odot })$ plot of high-redshift galaxies are shown in the upper panel of Fig \ref{fig:1}, and lower panel of Fig \ref{fig:1} illustrates  $\Delta (\log {\kern 1pt} SFR){\kern 1pt} $ against redshift, which implies that the slope $\gamma$ is dependent on the redshift.

\subsection{Filter segmented data for measuring luminosity distance}

As can be seen in the lower panel of Fig \ref{fig:1}, the redshift dependence of the slope $\gamma$ is indicated for the sample galaxies, so we fit our data points by segment with $4 < z \le 7$, $z >  7$, and $z \le 4$ to avoid the redshift dependence for the slope $\gamma$.  We adopt this segmented sample to investigate the $SFR - {M_ * }$ relation and calculate their luminosity distance.

\subsection{Parametric formula for SFR and stellar mass ${M_ * }$}

 Using relation $L = 4\pi {D_L}^2f$ in equation (\ref{eq3}), we get
\begin{eqnarray}\label{eq4}
\begin{array}{l}
\log SFR' = \Phi ({\frac{{{M_*}}}{{{M_ \odot }}}^\prime },D_L)\\
{\kern 1pt} {\kern 1pt} {\kern 1pt} {\kern 1pt} {\kern 1pt} {\kern 1pt} {\kern 1pt} {\kern 1pt} {\kern 1pt} {\kern 1pt} {\kern 1pt} {\kern 1pt} {\kern 1pt} {\kern 1pt} {\kern 1pt} {\kern 1pt} {\kern 1pt} {\kern 1pt} {\kern 1pt} {\kern 1pt} {\kern 1pt} {\kern 1pt} {\kern 1pt} {\kern 1pt} {\kern 1pt} {\kern 1pt} {\kern 1pt} {\kern 1pt} {\kern 1pt} {\kern 1pt} {\kern 1pt} {\kern 1pt} {\kern 1pt} {\kern 1pt} {\kern 1pt} {\kern 1pt} {\kern 1pt} {\kern 1pt} {\kern 1pt} = \beta  + \gamma \log ({\frac{{{M_*}}}{{{M_ \odot }}}^\prime }) + (\gamma  - 1)\log (4\pi D_L^2),
\end{array}
\end{eqnarray}
where $ SFR' = 4.6 \times {10^{ - 42}}\frac{{{F_{H\alpha }}}}{{erg{\kern 1pt} {\kern 1pt} {s^{ - 1}}}}[{M_ \odot }{\kern 1pt} y{r^{ - 1}}]$, and $SFR' = 1.4 \times {10^{ - 28}}\frac{{{F_{UV }}}}{{erg{\kern 1pt} {\kern 1pt} {s^{ - 1}}}}[{M_ \odot }{\kern 1pt} y{r^{ - 1}}]$ for the galaxies at $1.8 < z < 3.5$ and $5 < z < 14$ respectively. So $\log SFR =\log SFR' +\log (4\pi D_L^2)$. Similarly, $\log (\frac{{{M_*}}}{{{M_ \odot }}}) = \log (\frac{{{M_*}}}{{{M_ \odot }{\kern 1pt} L}}) + \log {\kern 1pt} F + \log (4\pi D_L^2) = \log ({\frac{{{M_*}}}{{{M_ \odot }}}^\prime }) + \log (4\pi D_L^2)$, or $\log {\kern 1pt} {M_ * }/{L_\lambda } = \log {\kern 1pt} {M_ * }^\prime /{f_\lambda }$. This equation can be effectively used for testing the correlation between SFR and stellar mass for galaxies.


We fit the $SFR - {M_ * }$ relation to galaxies by minimizing a likelihood function consisting of a modified ${\chi ^2}$ function based on MCMC, allowing for an intrinsic dispersion $\delta$
\begin{large}
 \begin{eqnarray}\label{eq5}
\begin{array}{l}
 - 2\ln L = \sum\limits_{i = 1}^N {\left\{ {\frac{{{{[\log {{({SFR'})}_i} - \Phi {{({\frac{{{M_*}}}{{{M_ \odot }}}^\prime },D_L)}_i}]}^2}}}{{s_i^2}}} \right\}} \\
{\kern 1pt} {\kern 1pt} {\kern 1pt} {\kern 1pt} {\kern 1pt} {\kern 1pt} {\kern 1pt} {\kern 1pt} {\kern 1pt} {\kern 1pt} {\kern 1pt} {\kern 1pt} {\kern 1pt} {\kern 1pt} {\kern 1pt} {\kern 1pt} {\kern 1pt} {\kern 1pt} {\kern 1pt} {\kern 1pt} {\kern 1pt} {\kern 1pt} {\kern 1pt} {\kern 1pt} {\kern 1pt} {\kern 1pt} {\kern 1pt} {\kern 1pt} {\kern 1pt} {\kern 1pt} {\kern 1pt} {\kern 1pt} {\kern 1pt} {\kern 1pt} {\kern 1pt} {\kern 1pt} {\kern 1pt} {\kern 1pt}  + \sum\limits_{i = 1}^N {\ln (2\pi s_i^2)} ,
\end{array}
\end{eqnarray}
\end{large}
where $\Phi ({\frac{{{M_*}}}{{{M_ \odot }}}^\prime },{D_L})$ is given by equation (\ref{eq4}), and ${s_i}^2 = \sigma _i^2(\log SFR') + {\gamma ^2} \cdot \sigma _i^2(\log {\kern 1pt} {\kern 1pt} ({\frac{{{M_*}}}{{{M_ \odot }}}^\prime })) + {\delta ^2}$,  $\sigma _i(\log SFR')$ and $\sigma _i(\log {\kern 1pt} {\kern 1pt} ({\frac{{{M_*}}}{{{M_ \odot }}}^\prime }))$ indicate the statistical errors for $\log SFR'$, $\log {\kern 1pt} {\kern 1pt} ({\frac{{{M_*}}}{{{M_ \odot }}}^\prime })$. $\delta$ is the intrinsic dispersion \citep{Kim2011}, which can be fitted as a free parameter.

 The large positive and negative space curvature of the universe have not been obviously observed by observational data, and we approximately adopt a curvature of ${\Omega _k} = 0$,  a prior cosmological constant $\Lambda CDM$ model is assumed, then ${\Omega _\Lambda }{\rm{ = }}1{\rm{ - }}{\Omega _m}$ when ${\Omega _R} \ll {\Omega _m}$. In this case, the free parameters are $\beta$, $\gamma$ and the intrinsic dispersion $\delta$, and the cosmological parameters ${\Omega _m}$. We note that the Hubble constant ${H_0}$ is absorbed into the parameter $\beta$ when fitting equation (\ref{eq4}), without an independent determination of this parameter, so we fix ${H_0} = 70{\kern 1pt} {\kern 1pt} km{\kern 1pt} {\kern 1pt} {\kern 1pt} {s^{ - 1}}{\kern 1pt} Mp{c^{ - 1}}$ \citep{Reid2019, Aghanim2020}.

\begin{table*}[htpb]
\setlength{\tabcolsep}{2mm}
 \centering
  \caption{ Model fitting results for galaxies}
   \label{tab:1}
  \vspace{0.3cm}
  \begin{tabular}{@{}cccccc@{}}
   \hline
 Sample&$\beta $&${\gamma} $&$\delta $&${\Omega _m}$&$- 2In{L_{\max }}/N$\\

   $Galaxies (z < 2.5)$&-2.294$\pm$0.431    &0.313$\pm$0.055 & 0.281$\pm$0.041 &0.302$\pm$0.031  &7.9/27\\
   $Galaxies (2.5 \le z < 4)  $&-0.772$\pm$0.831    &0.189$\pm$0.098 & 0.334$\pm$0.046 &0.306$\pm$0.03 &22.4/24\\
   $Galaxies (z > 5, {M_{UV}} \le  - 19)$&-1.967$\pm$0.2    &0.332$\pm$0.023 & 0.244$\pm$0.014 &0.324$\pm$0.02 &51.4/227\\
   $Galaxies (z > 5, {M_{UV}} >  - 19)$&-1.63$\pm$0.34    &0.23$\pm$0.042 & 0.225$\pm$0.018 &0.313$\pm$0.027  &10.9/114\\
   \hline

\end{tabular}
\end{table*}

\begin{table*}
\centering
\caption{The properties of the 392 high-redshift galaxies, $DM$ are the distance modulus from a fit of the relation $SFR \propto {\kern 1pt} {({M_ * }/{M_ \odot })^\gamma }$ with $\Lambda CDM$ model, ${\sigma _{DM}}$ are their error.  Only five of the objects are listed.}
\label{tab:2}
\vspace{0.3cm}
\begin{tabular}{ccccccccc}
\hline
ID &$R.A.$&$Decl.$&$z$&${M_{UV}}$&${M_ * }^\prime  $&$SFR'$&$DM$&${\sigma _{DM}}$\\
&deg.&deg.&&mag&$\log {M_ \odot }/{m^2}$&$\log {M_ \odot }/(yr{\kern 1pt} {\kern 1pt}  \cdot {m^2})$\\
\hline
JADESGDS-30934	&53.149883	&-27.776503	&13.2	&$-18.3_{ - 0.1}^{ + 0.1}$&	$-48.68_{-0.4}^{+0.4}	$ &$-56.08_{-0.1}^{+0.1}$&	50.28&0.95\\
JADESGDS-33832	&53.18993	&-27.771496	&11.8	&$-18.1_{-0.1}^{+0.1}$	&$-48.67_{-0.2}^{+0.2}$&	$-55.97_{-0.1}^{+0.1}$&	49.92&	0.96\\
JADESGDS-14022	&53.144043	&-27.804459	&8.9	&$-18.5_{-0.1}^{+0.1}$&	$-48.39_{-0.1}^{+0.1}$&	$-55.69_{-0.1}^{+0.1}$	&49.23	&0.95\\
JADESGDS-35079	&53.181484	&-27.769501	&7.4	&$-18.5_{-0.1}^{+0.1}$	&$-48.51_{-0.1}^{+0.1}$&	$-55.51_{-0.1}^{+0.1}$&	48.53&	1.03\\
A2744-6031&	3.5769939	&-30.415525	&5.13	&$-18.8_{-0.1}^{+0.1}$&	$-46.74_{-0.1}^{+0.1}$&	
$-54.94_ {-0.1}^{+0.1}$&	48.08&	0.95\\

\hline

\end{tabular}

\end{table*}

\begin{table*}[htpb]
\footnotesize
\setlength{\tabcolsep}{0.5mm}
 \centering
  \caption{Fit results on model parameters for a combination of SNla and galaxies}
\label{tab:3}
  \vspace{0.3cm}
  \begin{tabular}{@{}ccccccccc@{}}
  \hline
&$Sample $&Parameters&${\Omega _m}$&${w_0}$&${w_\alpha }$&$\chi _{Total}^2$/N\\
   $\Lambda CDM$
&SN+Galaxies ($z < 4$)   &Mean&   0.274$\pm$0.007 &$-$&$-$ \\
&                        &Best fit&   0.273 &$-$&$-$ &1103.7/1099\\
&SN+Galaxies ($z < 9.5$) &Mean& 0.275$\pm$0.007 &$-$&$-$ \\
&                        &Best fit&   0.274 &$-$&$-$ &1353.1/1393\\
&SN+Galaxies ($z < 14$) &Mean&  0.277$\pm$0.008 &$-$&$-$ \\
&                        &Best fit&   0.277 &$-$&$-$ &1411.6/1440\\

   \hline
      ${w_0}{w_a}CDM$
   &SN+Galaxies ($z < 4$) &Mean&   0.335$\pm$0.032 &-1.152$\pm$0.048&-0.209$\pm$0.754 \\
   &                        &Best fit&   0.331 &-1.162&0.118 &1099.6/1099\\
&SN+Galaxies ($z < 9.5$) &Mean&  0.286$\pm$0.028 &-1.085$\pm$0.048&0.391$\pm$0.476\\
&                        &Best fit&   0.286 &-1.096&0.534 &1347.9/1393\\
&SN+Galaxies ($z < 14$) &Mean& 0.329$\pm$0.033&-1.17$\pm$0.065&0.097$\pm$0.676\\
&                        &Best fit&   0.321 &-1.143&0.163 &1405.4/1440\\

   \hline

\end{tabular}
\end{table*}

Meanwhile, we measure the distance modulus for galaxies based on $SFR - {M_ * }$ relation. Distance modulus is defined as $DM = 5\log {\kern 1pt} ({D_L}) + 25$. Equation (\ref{eq4}) gives $DM$ as

 \begin{equation}\label{eq6}
{\rm{DM  = }}\frac{{5[\log SFR' - \gamma \log ({\frac{{{M_*}}}{{{M_ \odot }}}^\prime }) - \beta ']}}{{2(\gamma-1) }}+25,
\end{equation}
where $\beta ' = \beta  + (\gamma-1) \log {\kern 1pt} (4\pi )$.  The error is
 \begin{equation}\label{eq7}
{\sigma _{DM}} = DM\sqrt {{{(\frac{{{\sigma _f}}}{f})}^2} + {{(\frac{{{\sigma _\gamma }}}{\gamma-1 })}^2}} ,
\end{equation}
where $f=\log SFR' - \gamma \log ({\frac{{{M_*}}}{{{M_ \odot }}}^\prime }) - \beta '$, and ${\sigma _f}^2 = {\sigma ^2}(\log {\kern 1pt} SFR') + {\gamma ^2} \cdot {\sigma ^2}(\log {\kern 1pt} {\kern 1pt} ({\frac{{{M_*}}}{{{M_ \odot }}}^\prime }))+{\sigma _{\beta '}}^2$. From equation (\ref{eq7}),  the uncertainty of the slope $\gamma$ obviously influences the error of distance modulus for galaxies.

\subsection{Fitting result for the relation of $SFR - {M_ * }$}

We fit the equation (\ref{eq5}) to galaxies in segment by redshift and the absolute UV magnitude ${M_{UV}}$, the fitting results are shown in table \ref{tab:1}. The linear regression slope for all samples have $\gamma  > 0$, which suggests that the star formation rate is correlated with stellar mass ${M_ * }$. Meanwhile, the slope for galaxies at $z > 5{\kern 1pt} ,{\kern 1pt} {M_{UV}} \le  - 19$ and $z > 5{\kern 1pt} ,{\kern 1pt} {M_{UV}} >  - 19$ have $\gamma  = 0.332 \pm 0.023$ and $\gamma  = 0.23 \pm 0.042$, and there are also an obvious deference for $ - 2In{\kern 1pt} {\kern 1pt} {L_{\max }}/N$, which implies that the slope $\gamma$ is dependent on ${M_{UV}}$. The redshift dependence of $SFR - {M_ * }$ relation has also been found by many studies \citep{Dave2008, Koyama2013}, our results also support that there is a dependency upon the redshift for $SFR - {M_ * }$ from table \ref{tab:1}.

\begin{figure*}[htpb]
\centering
\includegraphics[width=\linewidth,scale=1.00]{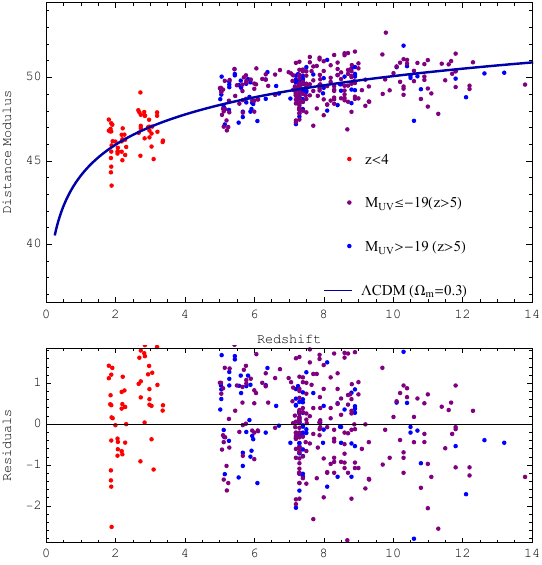}
\caption{The distance modulus of galaxies with $z < 4$, $z > 5{\kern 1pt}$ , and ${\kern 1pt} {M_{UV}} \le  - 19$, $ {M_{UV}} >  - 19$ from a fit of Equation (\ref{eq4}) when assuming $\Lambda CDM$ cosmology. The blue line shows a flat $\Lambda CDM$ model fit with ${\Omega _m} = 0.3$. The lower panel shows the residuals of the distance modulus at different redshifts}
\label{fig:2}
\end{figure*}

We obtain the distance modulus of galaxies by substituting the statistical average values of $\beta$ and $\gamma$ into equation (\ref{eq6}), which are shown in the upper panel of Fig \ref{fig:2}, including the galaxies with $z < 4$, $z > 5{\kern 1pt},{\kern 1pt} {M_{UV}} \le  - 19$ and $z > 5{\kern 1pt},{\kern 1pt} {M_{UV}} >  - 19$. Meanwhile the distance modulus and properties of the 392 galaxies are listed in Table \ref{tab:2}. The lower panel of Fig \ref{fig:2} shows the plot of the residuals of the distance modulus against redshift, the residuals are from the measuring distance modulus for galaxies and $\Lambda CDM$ cosmology $({\Omega _m} = 0.3)$. There could have several reasons for the large scatter in the luminosity distance, including observational error, and intrinsic variation of the star formation rate-stellar mass relation.

The distance modulus for high-redshift galaxies can be used to test the property of dark energy.

\section{The reconstruction of dark energy equation of state w(z)}\label{Sec:4}

Although the dark energy can be used to effectively explain the accelerating expansion of the universe and the cosmic microwave background (CMB) anisotropies \citep{Riess1998, Amanullah2010, Betoule2014, Scolnic2018, Conley2010, Aghanim2020, Hu2002, Spergel2003, Ade2016, Aghanim2016}, the
origin and nature of dark energy are still unclear.

The research methods of dark energy include two kinds. One is to try to explain the physical origin of its density and pressure by constraining dark energy physical models \citep{Peebles2003, Ratra1988, Li2004, Maziashvili2007, Amendola2000}. Understanding the physical nature of dark energy is very important for our universe. Whether or not the dark energy is composed of Fermion pairs in a vacuum or Boson pairs. The order of magnitude for the strength of dark energy is far smaller than that the elementary particles needed when they were created in the very early Universe. Another way is to investigate whether the density of dark energy changes over time, which can be checked by reconstructing the dark energy equation of state $w(z)$ \citep{Linder2003, Maor2002}, this is independent of physical models. High redshift observational data can better solve these problems.

 The reconstruction methods of the equation of state for dark energy can be classed into parametric and non-parametric methods \citep{Huterer2003, Clarkson2010, Holsclaw2010, Seikel2012, Shafieloo2012, Crittenden2012}. We apply the distance modulus of galaxies and SNla to reconstruct $w(z)$ by parametric method, which can be used for testing the property of dark energy.

SNla Pantheon sample is a combination of data sources of the Sloan Digital Sky Survey (SDSS), the Pan-STARRS1 (PS1), SNLS, and various low-z and Hubble Space Telescope samples. There are 335 SNIa presented by SDSS \citep{Betoule2014,Gunn2006,Gunn1998,Sako2007,Sako2018}, and PS1 provided 279 SNla \citep{Scolnic2018}. The rest of the Pantheon sample are from the ${\rm{CfA1 - 4}}$, CSP, and Hubble Space Telescope (HST) SN surveys \citep{Amanullah2010, Conley2010}. This joint sample of 1048 SNIa is called the Pantheon sample.

The integral formula of ${D_L} - z$ relation in near flat space is given by
 \begin{equation}\label{eq13}
\begin{array}{l}
{D_L} = \frac{{1 + z}}{{{H_0}}}\int_0^z {d{z'}[{\Omega _m}{{(1 + {z'})}^3}} \\
{\kern 1pt} {\kern 1pt} {\kern 1pt} {\kern 1pt} {\kern 1pt} {\kern 1pt} {\kern 1pt} {\kern 1pt} {\kern 1pt} {\kern 1pt} {\kern 1pt} {\kern 1pt} {\kern 1pt} {\kern 1pt} {\kern 1pt} {\kern 1pt} {\kern 1pt} {\kern 1pt} {\kern 1pt} {\kern 1pt} {\kern 1pt} {\kern 1pt} {\kern 1pt} {\kern 1pt} {\kern 1pt} {\kern 1pt} {\kern 1pt} {\kern 1pt} {\kern 1pt} {\kern 1pt} {\kern 1pt} {\kern 1pt} {\kern 1pt} {\kern 1pt} {\kern 1pt} {\kern 1pt} {\kern 1pt} {\kern 1pt} {\kern 1pt} {\kern 1pt} {\kern 1pt} {\kern 1pt} {\kern 1pt} {\kern 1pt} {\kern 1pt} {\kern 1pt} {\kern 1pt} {\kern 1pt} {\kern 1pt} {\kern 1pt} {\kern 1pt}  + {\Omega _R}{(1 + {z'})^4} + \Omega _{DE}^{(0)}{{\mathop{\rm e}\nolimits} ^{\int_0^{{z'}} {\frac{{1 + w({z^{''}})}}{{1 + {z^{''}}}}d{z^{''}}} }}{]^{ - 1/2}}
\end{array}
\end{equation}
where ${{\Omega _R}}$ is radiation density. ${\Omega _{DE}^{(0)}}$ is the present dark energy density and satisfies $\Omega _{DE}^{(0)} = 1 - {\Omega _m}$ when ignoring ${{\Omega _R}}$ \citep{Komatsu2011, Aghanim2020}, $w(z)$ is dark energy equation of state.
We choose ${w_0}{w_a}CDM$ model and the parametric form is
 \begin{equation}\label{eq13}
w(z) = {w_0} + {w_a}\frac{z}{{1 + z}}.
\end{equation}
Therefore dark energy density can be written as
 \begin{equation}\label{eq14}
{\Omega _{DE}}(z) = \Omega _{DE}^{(0)}{(1 + z)^{3(1 + {w_0} + {w_a})}}\exp [ - 3{w_a}z/(1 + z)].
\end{equation}

We fit ${{\rm{w}}_0}{{\rm{w}}_a}{\rm{CDM}}$ model parameters to Galaxies and SNla by minimizing $\chi _{Total}^2$, the $\chi _{Total}^2$ is
 \begin{eqnarray}\label{eq15}
\chi _{Total}^2 =  \chi _{Galaxies}^2 + \chi _{SN}^2,
\end{eqnarray}
where  $\chi^2$ can be expressed as
 \begin{equation}\label{eq16}
\chi^2 = \Delta {\mu ^T}C_{{\mu _{ob}}}^{ - 1}\Delta \mu ,
\end{equation}
where $\Delta \mu  = \mu  - {\mu _{th}}$. ${C_\mu }$ is the covariance matrix of the distance modulus $\mu $.

\begin{figure}[htpb]
\includegraphics[width=\linewidth]{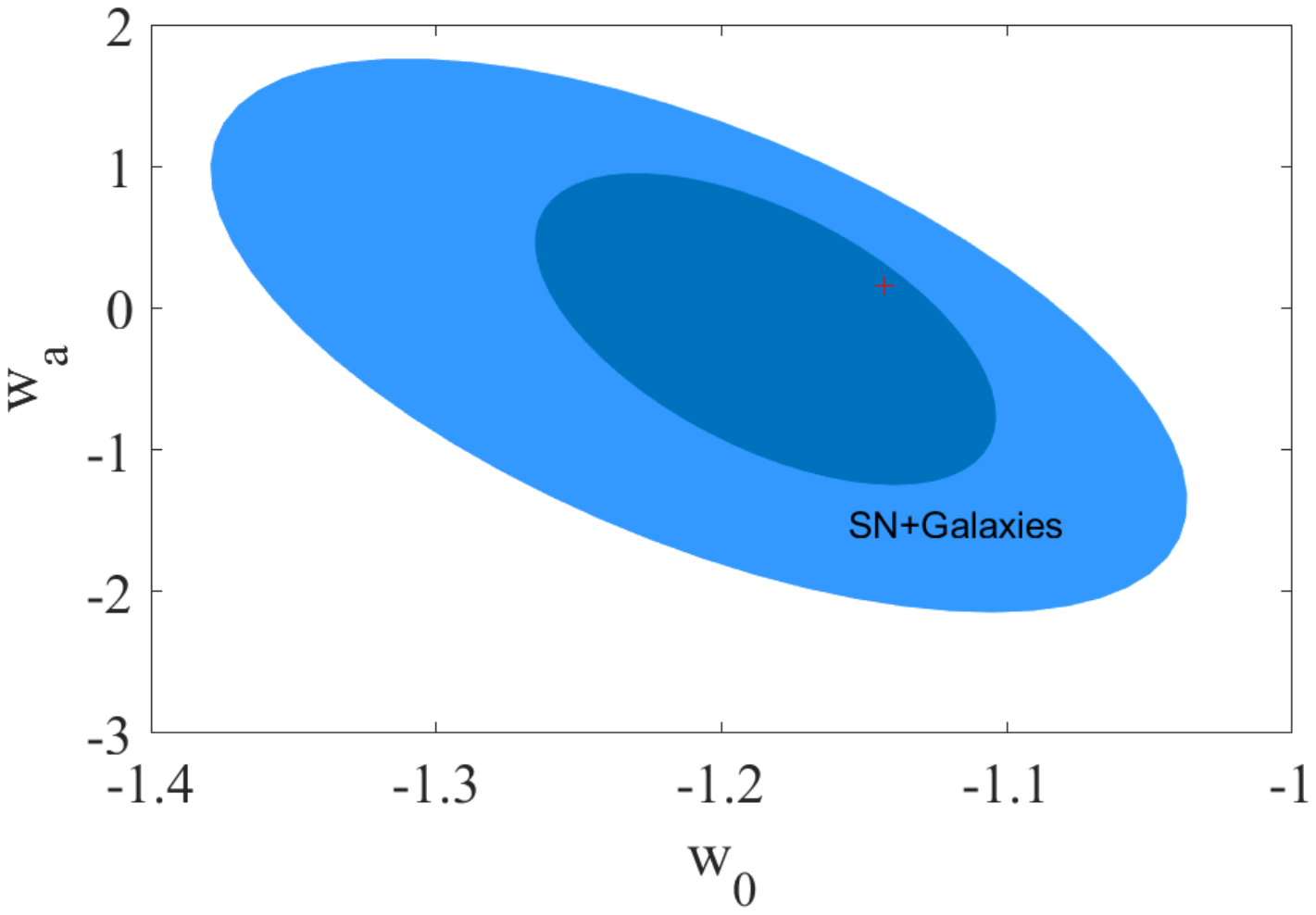}
\caption{$68\%$ and $95\%$ contours for ${w_0}$ and ${w_a}$ from a fit of the ${w_0}{w_a}CDM$ model to a combination of SNla and galaxies. The + dot in the responding color represents the best fitting values for ${w_0}$, ${w_a}$.}
\label{fig:3}
\end{figure}

We adopt equation (\ref{eq15}) to constrain model parameters, and fit results are illustrated in table \ref{tab:3}, ${w_0}{w_a}CDM$ has better goodness of fit than $\Lambda CDM$, and $\Delta \chi _{Total}^2$ is improved by $-6.2$, it implies $\Lambda CDM$ model is in tension with High-redshift galaxies at $\sim 2\sigma $, which is consistent with the results from the distance measurement using the X-ray luminosity relation of quasars \citep{Huang2022, Huang2024}. Meanwhile fig \ref{fig:3} shows  $68\%$ and $95\%$ contours for ${w_0}$ and ${w_a}$ from a fit of ${{\rm{w}}_0}{{\rm{w}}_a}{\rm{CDM}}$ model to a combination of SNla and galaxies.

\section{Summary}\label{Sec:5}

The investigation of the correlation for the star formation rate and stellar mass could make us understand more of the formation and evolution for galaxies. We obtain a sample of 392 galaxies from JWST by combining \citet{Li2023} and \citet{Morishita2024}. Firstly, we apply parametric methods to check the $SFR - {M_ * }$ relation, and get the slope $\gamma  > 0.15$,  which implies that the $SFR$ of galaxies is correlated with stellar mass ${M_ * }$.  We also test whether there is a dependency upon the redshift by residual analysis, the data suggest that the slope $\gamma$ depends on redshift.

Secondly, in order to avoid the absolute UV magnitude and redshift dependence for the slope $\gamma$, we divide the galaxies sample into discrete redshift bins, and fit the model to segmented data. The statistical results show that the slope $\gamma$ and $ - 2In{\kern 1pt} {\kern 1pt} {L_{\max }}/N$ from equation (\ref{eq5}) for galaxies with $z<4$ and $z>4$ have obvious differences, which indicates there is an obvious redshift dependence for $SFR \propto {\kern 1pt} {({M_ * }/{M_ \odot })^\gamma }$.

Finally, we obtain the luminosity distance of 392 galaxies from a fit of the star formation rate and stellar mass relation $SFR \propto {\kern 1pt} {({M_ * }/{M_ \odot })^\gamma }$ when assuming $\Lambda CDM$ cosmology, and use a joint of SNla and galaxies sample to reconstruct the dark energy equation of state $w(z)$ by parametric method and test the nature of dark energy. The data suggests ${w_0}{w_a}CDM$ model is superior to cosmological constant $\Lambda CDM$ model at $\sim 2\sigma $.

In the future, we will select more high redshift galaxies sample from the JWST NIRCam and NIRISS Extragalactic Fields, and hope to obtain galaxies at $z>7$ with the ${H\alpha }$ or ${H\beta }$ emission line and the continuum luminosity from future JWST MIRI. The high redshift observational data can be better used to reconstruct the equation of state and test the properties of dark energy, which involves whether or not the universe will keep expanding.

\section*{Acknowledgments}

  This work was supported by the National Science Foundation of China (grant No. 12303027), the Natural Science Foundation of Hebei Province (No. A2022109001).

\vspace{1cm}
\bibliographystyle{aasjournal}
\bibliography{bib}

\begin{thebibliography}{}
\expandafter\ifx\csname natexlab\endcsname\relax\def\natexlab#1{#1}\fi
\providecommand{\url}[1]{\href{#1}{#1}}
\providecommand{\dodoi}[1]{doi:~\href{http://doi.org/#1}{\nolinkurl{#1}}}
\providecommand{\doeprint}[1]{\href{http://ascl.net/#1}{\nolinkurl{http://ascl.net/#1}}}
\providecommand{\doarXiv}[1]{\href{https://arxiv.org/abs/#1}{\nolinkurl{https://arxiv.org/abs/#1}}}

\bibitem[{Ade {et~al.}(2016)Ade, Aghanim, Arnaud, Ashdown, Aumont, Baccigalupi,
  Banday, Barreiro, Bartlett, Bartolo, {et~al.}}]{Ade2016}
Ade, P., Aghanim, N., Arnaud, M., {et~al.} 2016, A\&A, 594, A24

\bibitem[{Aghanim {et~al.}(2016)Aghanim, Ashdown, Aumont, Baccigalupi,
  Ballardini, Banday, Barreiro, Bartolo, Basak, Battye, {et~al.}}]{Aghanim2016}
Aghanim, N., Ashdown, M., Aumont, J., {et~al.} 2016, A\&A, 596, A107

\bibitem[{Aghanim {et~al.}(2020)Aghanim, Akrami, Ashdown, Aumont, Baccigalupi,
  Ballardini, Banday, Barreiro, Bartolo, Basak, {et~al.}}]{Aghanim2020}
Aghanim, N., Akrami, Y., Ashdown, M., {et~al.} 2020, A\&A, 641, A6

\bibitem[{Ahumada {et~al.}(2020)Ahumada, Prieto, Almeida, Anders, Anderson,
  Andrews, Anguiano, Arcodia, Armengaud, Aubert, {et~al.}}]{Ahumada2020}
Ahumada, R., Prieto, C.~A., Almeida, A., {et~al.} 2020, APJS, 249, 3

\bibitem[{Amanullah {et~al.}(2010)Amanullah, Lidman, Rubin, Aldering, Astier,
  Barbary, Burns, Conley, Dawson, Deustua, {et~al.}}]{Amanullah2010}
Amanullah, R., Lidman, C., Rubin, D., {et~al.} 2010, APJ, 716, 712

\bibitem[{Amendola(2000)}]{Amendola2000}
Amendola, L. 2000, Phys. Rev. D, 62, 043511

\bibitem[{Bakx {et~al.}(2023)Bakx, Zavala, Mitsuhashi, Treu, Fontana, Tadaki,
  Casey, Castellano, Glazebrook, Hagimoto, {et~al.}}]{Bakx2023}
Bakx, T.~J., Zavala, J.~A., Mitsuhashi, I., {et~al.} 2023, MNRAS, 519, 5076

\bibitem[{Bergamini {et~al.}(2023)Bergamini, Acebron, Grillo, Rosati, Caminha,
  Mercurio, Vanzella, Mason, Treu, Angora, {et~al.}}]{Bergamini2023}
Bergamini, P., Acebron, A., Grillo, C., {et~al.} 2023, APJ, 952, 84

\bibitem[{Betoule {et~al.}(2014)Betoule, Kessler, Guy, Mosher, Hardin, Biswas,
  Astier, El-Hage, Konig, Kuhlmann, {et~al.}}]{Betoule2014}
Betoule, M., Kessler, R., Guy, J., {et~al.} 2014, A\&A, 568, A22

\bibitem[{Bouwens {et~al.}(2015)Bouwens, Illingworth, Oesch, Trenti, Labb{\'e},
  Bradley, Carollo, Van~Dokkum, Gonzalez, Holwerda, {et~al.}}]{Bouwens2015}
Bouwens, R.~J., Illingworth, G., Oesch, P., {et~al.} 2015, APJ, 803, 34

\bibitem[{Brinchmann {et~al.}(2004)Brinchmann, Charlot, White, Tremonti,
  Kauffmann, Heckman, \& Brinkmann}]{Brinchmann2004}
Brinchmann, J., Charlot, S., White, S.~D., {et~al.} 2004, MNRAS, 351, 1151

\bibitem[{Chabrier(2003)}]{Chabrier2003}
Chabrier, G. 2003, PASP, 115, 763

\bibitem[{Clarkson \& Zunckel(2010)}]{Clarkson2010}
Clarkson, C., \& Zunckel, C. 2010, Phys. Rev. Lett., 104, 211301

\bibitem[{Conley {et~al.}(2010)Conley, Guy, Sullivan, Regnault, Astier,
  Balland, Basa, Carlberg, Fouchez, Hardin, {et~al.}}]{Conley2010}
Conley, A., Guy, J., Sullivan, M., {et~al.} 2010, APJS, 192, 1

\bibitem[{Crittenden {et~al.}(2012)Crittenden, Zhao, Pogosian, Samushia, \&
  Zhang}]{Crittenden2012}
Crittenden, R.~G., Zhao, G.-B., Pogosian, L., Samushia, L., \& Zhang, X. 2012,
  J. Cosmol. Astropart. Phys., 2012, 048

\bibitem[{Daddi {et~al.}(2007)Daddi, Dickinson, Morrison, Chary, Cimatti,
  Elbaz, Frayer, Renzini, Pope, Alexander, {et~al.}}]{Daddi2007}
Daddi, E., Dickinson, M., Morrison, G., {et~al.} 2007, APJ, 670, 156

\bibitem[{Dav{\'e}(2008)}]{Dave2008}
Dav{\'e}, R. 2008, MNRAS, 385, 147

\bibitem[{Elbaz {et~al.}(2007)Elbaz, Daddi, Le~Borgne, Dickinson, Alexander,
  Chary, Starck, Brandt, Kitzbichler, MacDonald, {et~al.}}]{Elbaz2007}
Elbaz, D., Daddi, E., Le~Borgne, D., {et~al.} 2007, A\& A, 468, 33

\bibitem[{Golubchik {et~al.}(2022)Golubchik, Furtak, Meena, \&
  Zitrin}]{Golubchik2022}
Golubchik, M., Furtak, L.~J., Meena, A.~K., \& Zitrin, A. 2022, APJ, 938, 14

\bibitem[{Gunn {et~al.}(1998)Gunn, Carr, Rockosi, Sekiguchi, Berry, Elms,
  De~Haas, Ivezi{\'c}, Knapp, Lupton, {et~al.}}]{Gunn1998}
Gunn, J., Carr, M., Rockosi, C., {et~al.} 1998, AJ, 116, 3040

\bibitem[{Gunn {et~al.}(2006)Gunn, Siegmund, Mannery, Owen, Hull, Leger, Carey,
  Knapp, York, Boroski, {et~al.}}]{Gunn2006}
Gunn, J.~E., Siegmund, W.~A., Mannery, E.~J., {et~al.} 2006, AJ, 131, 2332

\bibitem[{Holsclaw {et~al.}(2010)Holsclaw, Alam, Sanso, Lee, Heitmann, Habib,
  \& Higdon}]{Holsclaw2010}
Holsclaw, T., Alam, U., Sanso, B., {et~al.} 2010, Phys. Rev. Lett., 105, 241302

\bibitem[{Hu \& Dodelson(2002)}]{Hu2002}
Hu, W., \& Dodelson, S. 2002, ARA\&A, 40, 171

\bibitem[{Huang \& Chang(2022)}]{Huang2022}
Huang, L., \& Chang, Z. 2022, MNRAS, 515, 1358

\bibitem[{Huang {et~al.}(2024)Huang, Tu, Chang, Song, He, \& Fu}]{Huang2024}
Huang, L., Tu, Z., Chang, N., {et~al.} 2024, Phys. Rev. D, 109, 043529

\bibitem[{Huterer \& Starkman(2003)}]{Huterer2003}
Huterer, D., \& Starkman, G. 2003, Phys. Rev. Lett., 90, 031301

\bibitem[{Kennicutt(1998)}]{Kennicutt1998}
Kennicutt, R.~C. 1998, APJ, 498, 541

\bibitem[{Kim(2011)}]{Kim2011}
Kim, A.~G. 2011, PASP, 123, 230

\bibitem[{Komatsu {et~al.}(2011)Komatsu, Smith, Dunkley, Bennett, Gold,
  Hinshaw, Jarosik, Larson, Nolta, \& Page}]{Komatsu2011}
Komatsu, E., Smith, K.~M., Dunkley, J., {et~al.} 2011, APJS, 192, 18

\bibitem[{Koyama {et~al.}(2013)Koyama, Smail, Kurk, Geach, Sobral, Kodama,
  Nakata, Swinbank, Best, Hayashi, {et~al.}}]{Koyama2013}
Koyama, Y., Smail, I., Kurk, J., {et~al.} 2013, MNRAS, 434, 423

\bibitem[{Li(2004)}]{Li2004}
Li, M. 2004, Phys. Lett. B, 603, 1

\bibitem[{Li {et~al.}(2023)Li, Cai, Bian, Lin, Li, Wu, Sun, Zhang, Golden-Marx,
  Sun, {et~al.}}]{Li2023}
Li, M., Cai, Z., Bian, F., {et~al.} 2023, APJL, 955, L18

\bibitem[{Linder(2003)}]{Linder2003}
Linder, E.~V. 2003, Phys. Rev. Lett., 90, 091301

\bibitem[{Lyke {et~al.}(2020)Lyke, Higley, McLane, Schurhammer, Myers, Ross,
  Dawson, Chabanier, Martini, Des~Bourboux, {et~al.}}]{Lyke2020}
Lyke, B.~W., Higley, A.~N., McLane, J., {et~al.} 2020, APJS, 250, 8

\bibitem[{Mahler {et~al.}(2018)Mahler, Richard, Cl{\'e}ment, Lagattuta,
  Schmidt, Patr{\'\i}cio, Soucail, Bacon, Pello, Bouwens,
  {et~al.}}]{Mahler2018}
Mahler, G., Richard, J., Cl{\'e}ment, B., {et~al.} 2018, MNRAS, 473, 663

\bibitem[{Maor {et~al.}(2002)Maor, Brustein, McMahon, \& Steinhardt}]{Maor2002}
Maor, I., Brustein, R., McMahon, J., \& Steinhardt, P.~J. 2002, Phys. Rev. D,
  65, 123003

\bibitem[{Maziashvili(2007)}]{Maziashvili2007}
Maziashvili, M. 2007, Int. J. Mod. Phys. D, 16, 1531

\bibitem[{Morishita(2021)}]{Morishita2021}
Morishita, T. 2021, APJS, 253, 4

\bibitem[{Morishita {et~al.}(2024)Morishita, Stiavelli, Chary, Trenti,
  Bergamini, Chiaberge, Leethochawalit, Roberts-Borsani, Shen, \&
  Treu}]{Morishita2024}
Morishita, T., Stiavelli, M., Chary, R.-R., {et~al.} 2024, APJ, 963, 9

\bibitem[{Naidu {et~al.}(2022)Naidu, Oesch, van Dokkum, Nelson, Suess, Brammer,
  Whitaker, Illingworth, Bouwens, Tacchella, {et~al.}}]{naidu2022two}
Naidu, R.~P., Oesch, P.~A., van Dokkum, P., {et~al.} 2022, APJL, 940, L14

\bibitem[{Noeske {et~al.}(2007)Noeske, Weiner, Faber, Papovich, Koo,
  Somerville, Bundy, Conselice, Newman, Schiminovich, {et~al.}}]{Noeske2007}
Noeske, K., Weiner, B., Faber, S., {et~al.} 2007, APJ, 660, L43

\bibitem[{Peebles \& Ratra(2003)}]{Peebles2003}
Peebles, P. J.~E., \& Ratra, B. 2003, Rev. Mod. Phys, 75, 559

\bibitem[{Ratra \& Peebles(1988)}]{Ratra1988}
Ratra, B., \& Peebles, P.~J. 1988, Phys. Rev. D, 37, 3406

\bibitem[{Reid {et~al.}(2019)Reid, Pesce, \& Riess}]{Reid2019}
Reid, M., Pesce, D.~W., \& Riess, A. 2019, APJL, 886, L27

\bibitem[{Riess {et~al.}(1998)Riess, Filippenko, Challis, Clocchiatti, Diercks,
  Garnavich, Gilliland, Hogan, Jha, Kirshner, {et~al.}}]{Riess1998}
Riess, A.~G., Filippenko, A.~V., Challis, P., {et~al.} 1998, AJ, 116, 1009

\bibitem[{Sako {et~al.}(2007)Sako, Bassett, Becker, Cinabro, DeJongh, Depoy,
  Dilday, Doi, Frieman, Garnavich, {et~al.}}]{Sako2007}
Sako, M., Bassett, B., Becker, A., {et~al.} 2007, AJ, 135, 348

\bibitem[{Sako {et~al.}(2018)Sako, Bassett, Becker, Brown, Campbell, Wolf,
  Cinabro, D’andrea, Dawson, DeJongh, {et~al.}}]{Sako2018}
Sako, M., Bassett, B., Becker, A.~C., {et~al.} 2018, Publ. Astron. Soc. Aust.,
  130, 064002

\bibitem[{Scolnic {et~al.}(2018)Scolnic, Jones, Rest, Pan, Chornock, Foley,
  Huber, Kessler, Narayan, Riess, {et~al.}}]{Scolnic2018}
Scolnic, D.~M., Jones, D., Rest, A., {et~al.} 2018, APJ, 859, 101

\bibitem[{Seikel {et~al.}(2012)Seikel, Clarkson, \& Smith}]{Seikel2012}
Seikel, M., Clarkson, C., \& Smith, M. 2012, J. Cosmol. Astropart. Phys., 2012,
  036

\bibitem[{Shafieloo {et~al.}(2012)Shafieloo, Kim, \& Linder}]{Shafieloo2012}
Shafieloo, A., Kim, A.~G., \& Linder, E.~V. 2012, Phys. Rev. D, 85, 123530

\bibitem[{Skelton {et~al.}(2014)Skelton, Whitaker, Momcheva, Brammer,
  Van~Dokkum, Labb{\'e}, Franx, Van Der~Wel, Bezanson, Da~Cunha,
  {et~al.}}]{Skelton2014}
Skelton, R.~E., Whitaker, K.~E., Momcheva, I.~G., {et~al.} 2014, APJS, 214, 24

\bibitem[{Spergel {et~al.}(2003)Spergel, Verde, Peiris, Komatsu, Nolta,
  Bennett, Halpern, Hinshaw, Jarosik, Kogut, {et~al.}}]{Spergel2003}
Spergel, D.~N., Verde, L., Peiris, H.~V., {et~al.} 2003, APJS, 148, 175

\bibitem[{Wambsganss(1998)}]{Wambsganss1998}
Wambsganss, J. 1998, Living Reviews in Relativity, 1, 1

\bibitem[{Whitaker {et~al.}(2012)Whitaker, Van~Dokkum, Brammer, \&
  Franx}]{Whitaker2012}
Whitaker, K.~E., Van~Dokkum, P.~G., Brammer, G., \& Franx, M. 2012, APJL, 754,
  L29

\end{thebibliography}


\end{document}